\documentclass[11pt]{article}
\usepackage{amssymb, amsmath, stmaryrd}
\usepackage{graphicx}
\newcommand{\beq}[1]{\begin{equation}\label{#1}}
\newcommand{\eeq}{\end{equation}}
\newcommand{\req}[1]{(\ref{#1})}
\newcommand{\bmu}[1]{\begin{multline}\label{#1}}
\newcommand{\emu}{\end{multline}}

\renewcommand{\(}{\left(}
\renewcommand{\)}{\right)}
\renewcommand{\[}{\left[}
\renewcommand{\]}{\right]}

\newcommand{\eq}{\triangleq}
\newcommand{\bigtri}{\bigtriangledown}

\renewcommand{\S}{{\mathcal S}}
\renewcommand{\L}{{\mathcal L}}
\newcommand{\N}{{\mathcal N}}
\newcommand{\Tau}{{\mathcal T}}
\newcommand{\x}{{\textbf{\textit{x}}}}
\newcommand{\y}{{\textbf{\textit{y}}}}
\newcommand{\z}{{\textbf{\textit{z}}}}
\renewcommand{\u}{{\textbf{\textit{u}}}}
\renewcommand{\v}{{\textbf{\textit{v}}}}
\renewcommand{\a}{{\textbf{\textit{a}}}}

\newcommand{\0}{{\textbf{\textit{0}}}}
\newcommand{\1}{{\textbf{\textit{1}}}}

\begin{document}

\begin{center}
{\Large\bf Symmetric Disjunctive List-Decoding Codes}
\\[15pt]
{\bf A.G. D'yachkov, \quad I.V. Vorobyev, \quad N.A. Polyanskii,\quad V.Yu. Shchukin}
\\[15pt]
Moscow State University, Faculty of Mechanics and Mathematics,\\
Department of Probability Theory, Moscow, 119992, Russia,\\
{\sf agd-msu@yandex.ru,\quad vorobyev.i.v@yandex.ru,\quad nikitapolyansky@gmail.com,\quad vpike@mail.ru}
\end{center}

\textbf{Abstract.}\quad
In this paper, we consider {\em symmetric disjunctive list-decoding} (SLD) codes, which are
a class of binary codes
based on a {\em symmetric disjunctive sum} (SDS) of binary symbols. By definition,
the SDS takes values from the ternary alphabet $\{0, 1, *\}$, where
the  symbol~$*$ denotes ``erasure''. Namely:  SDS  is equal to $0$ ($1$) if all its binary symbols
are equal to $0$ ($1$), otherwise SDS is equal to~$*$. The main purpose of this work is
to obtain bounds on the rate of these codes.
\medskip


\section{Statement of Problem and Results}
\subsection{Notations and Definitions}
\indent \indent
Let $N$, $t$, $s$, and $L$ be integers, where $2 \le s < t$, $1 \le L \le t - s$.
Let $\eq$ denote the equality by definition, $|A|$ -- the size of the set $A$ and
$[N] \eq \{1, 2, \dots, N\}$ - the set of integers from $1$ to $N$. The standard symbol
$\lfloor a \rfloor $ will be used to denote the largest integer $\le a$.

A binary $(N \times t)$-matrix
$$
X = \| x_i(j) \|, \quad x_i(j) = 0, 1, \quad \x_i \eq (x_i(1), \dots, x_i(t)), \quad \x(j) \eq (x_1(j), \dots, x_N(j)),
$$
$ i \in [N] $, $ j \in [t] $, with $N$ rows $\x_1, \dots, \x_N$ and $t$ columns $\x(1), \dots, \x(t)$ (codewords)
is called a {\em binary code of length $N$  and size $t = \lfloor 2^{RN} \rfloor$},
where a fixed parameter $R > 0$ is called a {\em rate} of the code~$X$.
The number of $1$'s in the codeword $x(j)$, i.e., $|\x(j)| \eq \sum\limits_{i = 1}^N \, x_i(j)$, is called a {\em weight} of $x(j)$, $j \in [t]$. A code $X$ is called a {\em constant weight binary code of  weight $w$}, $1 \le w < N$, if for any $j \in [t]$, the weight $|\x(j)| = w$.

Let $\u \bigvee \v$ denote the disjunctive sum of  binary columns $\u,\v\in\{0,1\}^N$.
%and $\v\in\{0,1\}^N$.
If  $\x,\y\in\{0, 1, *\}^N$ are arbitrary {\em ternary} columns  with components from
the alphabet $\{0, 1, *\}$,
then
%their {\em symmetric disjunctive sum}~\cite{s71} $\z = \x \bigtri \y$  is
the ternary column $\z=(z_1,z_2,\dots,z_N)\in\{0, 1, *\}^N$,
%$\z\in\{0, 1, *\}^N$ defined as follows
$$
%\forall \, i \in [N]
\qquad z_i \eq
\begin{cases}
0, \quad &\text{if} \quad x_i = y_i = 0,\\
1, \quad &\text{if} \quad x_i = y_i = 1,\\
*, \quad &\text{otherwise},
\end{cases}
$$
is called a {\em symmetric disjunctive sum}~\cite{s71} of $\x$ and $\y$.
This operation will be  denoted by $\bigtri$, that is $\z = \x \bigtri \y$. We say that a binary column $\u$ {\em covers}
a column $\v$ ($\u \succeq \v$) if $\u \bigvee \v = \u$, and a ternary column $\u$ {\em symmetrically covers} a column
$\v$ ($\u \trianglerighteq \v$) if $\u \bigtri \v = \u$.

\subsection{Symmetric Disjunctive List-Decoding Codes (SLD $s_L$-codes)}
\indent \indent
\textbf{Definition 1.}~\cite{dr83,ks64}.\quad
A binary code $X$ is said to be a {\em disjunctive list-decoding code of strength $s$ with list size}
$L$ (LD $s_L$-code) if the disjunctive sum of any $s$ codewords of $X$ covers not more than $L-1$ other codewords
of $X$ that are not components of the given sum. In other words, for any two disjoint sets
$\S, \L \subset [t], |\S| = s, |\L| = L, \S \cap \L = \varnothing$, there exist a row $\x_i, i \in [N]$,
and a column $\x(j), j \in \L$, such that
$$
x_i(k) = 0 \quad \forall k \in \S \qquad \text{and} \qquad x_i(j) = 1.
$$
Denote by $t_{ld}(N,s,L)$ the maximal size of LD $s_L$-codes of length $N$ and by $N_{ld}(t,s,L)$ the minimal length of LD $s_L$-codes of size~$t$. Define the {\em rate} of LD $s_L$-codes:
\beq{RsL}
R_L(s) \eq \varlimsup_{N \to \infty} \frac{\log_2 t_{ld}(N,s,L)}{N} \, = \, \varlimsup_{t\to\infty} \frac{\log_2 t}{N_{ld}(t,s,L)}.
\eeq

\textbf{Definition 2.}~\cite{dr81,d98,d03}.\quad
A binary code $X$ is said to be a {\em symmetric disjunctive list-decoding code of strength $s$ with list size $L$} (SLD $s_L$-code)
if the symmetric disjunctive sum of any $s$ codewords of $X$ symmetrically covers not more than $L-1$ other codewords of $X$
that are not components of the given sum. In other words, for any two disjoint sets
$\S, \L \subset [t]$, $|\S| = s$, $|\L| = L$, $\S \cap \L = \varnothing$, there exist a row $\x_i, i \in [N]$,
and a column $\x(j), j \in \L$, such that
\begin{align*}
&x_i(k) = 0 \quad \forall k \in \S \qquad \text{and} \qquad x_i(j) = 1, \quad \text{or}\\
&x_i(k) = 1 \quad \forall k \in \S \qquad \text{and} \qquad x_i(j) = 0.
\end{align*}
Denote by $t_{sld}(N,s,L)$ the maximal size of SLD $s_L$-codes of length $N$ and by $N_{sld}(t,s,L)$
the minimal length of SLD $s_L$-codes of size~$t$. Define the {\em rate} of SLD $s_L$-codes:
\beq{R*sL}
R^*_L(s) \eq \varlimsup_{N \to \infty} \frac{\log_2 t_{sld}(N,s,L)}{N} \, = \, \varlimsup_{t\to\infty} \frac{\log_2 t}{N_{sld}(t,s,L)}.
\eeq

\textbf{Remark 1.}\quad
An SLD $s_1$-code is the special case of separating codes~\cite{fgu69}. More specifically,
for $L = 1$, Definition~2 coincides with the definition of $(s, 1)$-separating code with
the alphabet size $q = 2$. Some results and applications of $(s, 1)$-separating binary codes
are presented in the survey~\cite{cs03}.

\textbf{Theorem 1.}\quad (Monotonicity properties).\quad
{\em The rate of SLD $s_L$-codes satisfies the following inequalities}
\beq{monoton}
R^*_L(s+1) \le R^*_L(s) \le R^*_{L+1}(s).
\eeq

\textbf{Proof of Theorem 1.}\quad
It immediately follows from Definition~2 that every SLD $(s+1)_L$-code is the corresponding SLD $s_L$-code,
so the left inequality in~\req{monoton} takes place.
Simultaneously, every SLD $s_L$-code is SLD $s_{L+1}$-code,
therefore the right inequality in~\req{monoton} is true.
$\square$

\subsection{Applications of Symmetric Disjunctive Codes}
\indent \indent
Applications of SLD $s_L$-codes relate to the {\em non-adaptive symmetric group testing} which is
based on the symmetric disjunctive sum of binary symbols\footnote{The adaptive symmetric group 
testing  for the search of  binomial sample was considered in~\cite{s71}.}.
Group testing deals with identification of defective units in a given pool. We use symmetric group tests, i.e.,
take a subset of the pool and check it.
The outcome of a symmetric group test belongs to the ternary alphabet. It is equal to~$0$, $1$ or~$*$,
if all tested units are not defective, all units are defective or at least one unit is defective and 
at least  another one is not defective, respectively.
The symmetric group testing was motivated by applications~\cite{s71} in electrical devices testing~(a) and chemical analysis~(b).

(a). Consider the situation, where one need to test electrical devices  such  as  conductors  (not  light  bulbs  that  give  a
visual  result)~\cite{s71}.  These  conductors  are  connected  both  in  parallel  and  in  series  and  the  results  for  these  two  
arrangements  are  obtained  separately  by  throwing  a  switch.  If  we get  current  for  the  series  configuration  then  all  are  good.
If  we  get  no  current  for  the  parallel  configuration  then  all are  defective.  In  the  one  remaining  case  (no  current  for
the  series  configuration  and  current  for  the  parallel  configuration),  we  have  at  least  one good  unit  and  at  least  $1$  
defective  unit.  Hence  for  our  purposes,  this  compound  test  to determine  which  of  these  three  situations  holds  is  
to  be  regarded  as  a  single  test  and  we  wish  to  minimize  the  number  of  such  tests.

%Consider the situation, where one need to test electrical devices such as conductors~\cite{s71}.
%These conductors are connected both in parallel and in series. The type of connection can be changed by a switch.
%If we get no current for the parallel connection, then all tested conductors are defective, and the outcome of test equals $1$.
%If we get current for the series connection, then all tested conductors are not defective, and the outcome is equal to $0$.
%Otherwise, if there is current for the parallel connection and there is no current for the series connection, the outcome equals $*$.
%This compound test is considered to be a single test, and our purpose is to detect all defective conductors in the smallest  number of tests.
%\medskip

(b). The second  possible  application  is  in  the  chemical  analysis of  several  specimens~\cite{s71},  
where  it  is  known a  priori  that  each specimen  contains  either  $A$  or  $B$  but  not  both,  which  are
two  specific  substances  of  interest.  Suppose  a  mixture  of several  specimens  is  formed and  then  we  split  
the  result  into  $2$  {\em aliquot  parts}.  By  using  reagent  $\alpha$,  which  precipitates  $A$  and  does  not  react  with  $B$, 
we  can  detect  ``no  $A$'' by  no  precipitate  in  one  of  the  $2$  aliquot  parts.  Similarly,  by using  reagent  $\beta$,  
which  precipitates  $B$  and  does  not  react with  $A$,  we  can  detect  ``no  $B$''  by  no  precipitate  in  the
other  of  the  $2$  aliquot  parts.  ``Some $A$ and some  $B$''  is  indicated  if both  reagents  cause  precipitation.  
Regarding  this  compound test  as  a  single  test,  we  want  to  classify  the  specimens  as
containing  $A$  or  containing  $B$  in  the  smallest  number  of  tests.

%The second illustrative example of symmetric group testing is the chemical component analysis of several specimens~\cite{s71}.
%Each specimen contains $A$ or $B$ but not both. Assume a mixture of some specimens is formed and separated into two parts.
%The use of reagent $\alpha$ ($\beta$), which indicates the substance $A$ ($B$) and does not react with $B$ ($A$),
%allows to detect the event ``No $A$'' (``No $B$''), if the reaction did not occur. The event ``$A$ and $B$'' is indicated if both reactions occur.
%This compound test is regarded as a single test. We wish to classify the specimens as containing $A$ or containing $B$ and to minimize the number of tests.

Suppose the size of the pool equals $t$ and the number of defected units does not exceed $s$.
As is the case with LD $s_L$-codes~\cite{v98}, SLD $s_L$-codes can be considered
in connection with the problem
of constructing \textit{two-stage non-adaptive symmetric group testing procedures}.
In the first stage, one does $N$ tests that can be depicted as an binary $(N \times t)$-matrix $X = \| x_i(j) \|$,
where a column $\x(j)$ corresponds to the $j$-th unit, a row $\x_i$ corresponds to the $i$-th test and
$x_i(j) \eq 1$ if and only if the $j$-th unit is included into the $i$-th testing group.
Then the ternary column $y$ of the test results equals the symmetric disjunctive sum of the columns which correspond to
the defective units. Let $X$ be SLD $s_L$-code, after decoding of the result column $y$, i.e. search of codewords which
are symmetrically covered by $y$, a set of $\le s + L - 1$ elements is selected. These units are
separately tested in the second stage. Note that for $s \ge 2$ the rate $R^*_L(s)$ of SLD $s_L$-codes
is a monotonically nondecreasing function of $L \ge 1$, and its limit
$$
R^*_{\infty}(s) = \lim_{L \to \infty} R^*_L(s)
$$
can be interpreted as the \textit{maximum rate} of two-stage non-adaptive symmetric group testing procedures in a search for $\le s$ defects
with the use of SLD $s_L$-codes.

In papers~\cite{dr81,d98}, we suggested another application of SLD codes called \textit{reference communication system}.
Let a system contain $M$ \textit{terminal stations} $\text{S}_1, \text{S}_2, ..., \text{S}_M$ and let
a \textit{multiple-access channel} (MAC) connect these $M$ stations to a \textit{central station} (CS).
Each terminal station has a \textit{source}. In every time interval, the source can produce a binary \textit{information packet}
of length $K$. Introduce $t \eq 2^K$ and enumerate all $2^K$ possible information packets by integers from $1$ to $t$.
The packets are encoded into binary sequences of length~$N$ by a code $X = (\x(i), i \in [t])$, where the codeword $\x(i), i \in [t],$ is the encoded sequence corresponding to
the information packet number $i$. Denote by $\S$ the set of numbers of generated packets and suppose $|\S| \le s$.

\begin{figure}[h!]
\centering{\includegraphics[width=140mm]{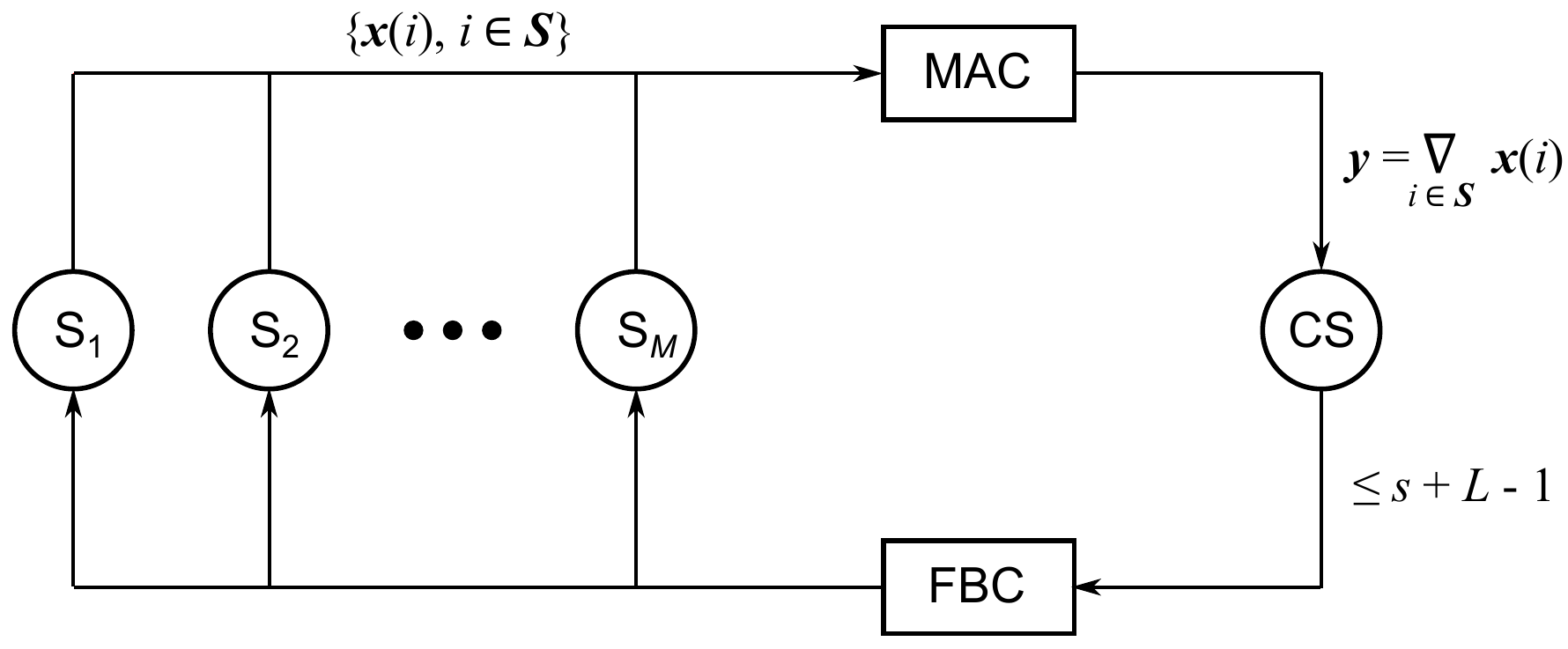}}
\caption{Reference communication system}
\label{PictureRCS}
\end{figure}

The CS is interested only in the contents of the received packet and not in the senders. Using a \textit{feedback broadcast channel} (FBC) the CS answers all $M$ stations to all requests. The model of MAC corresponds to the \textit{frequency modulation}, i.e., the output ternary sequence $\y$ is the symmetric disjunctive sum of the inputs. The scheme of reference communication system is represented on Figure~\ref{PictureRCS}.

Let the terminal stations use an SLD $s_L$-code $X$. 
Since the number of information packets produced by the terminal stations in the same time interval is not more than $s$, 
the CS is able to recover at most $s + L - 1$ packets, which contain $s$ transmitted packets.

Note that the model of MAC can also correspond to the \textit{impulse modulation}, i.e., the output binary sequence 
is the disjunctive sum of the inputs. In this case, it is convenient to use LD $s_L$-codes for encoding and decoding 
information packets. The case of impulse modulation was considered in~\cite{d98}.% and recently in~\cite{d14ad}.

Another application of SLD $s_1$-codes concerns with \textit{undetermined data}~\cite{sholomov13ran,sholomov13}.
Given an alphabet $A = \{a_1, a_2, ..., a_t\}$ of \textit{basic symbols}, to every nonempty subset $T \subseteq [t]$,
assign a symbol $a_T$, which is called \textit{undetermined}. Its \textit{specification} is any basic symbol
$a_i, i \in T$. By a \textit{specification} of a sequence of undetermined symbols we mean the result
of replacing all its symbols by some of its specifications.
The symbol $a_{[t]}$ that can be specified by any basic symbol is called \textit{indefinite} and
is denoted by $*$. Let $\Tau$ be a system of subsets $T \subseteq [t]$ and let $A^* = A^*_{\Tau} = \{ a_T | \; T \in \Tau \}$
be an \textit{undetermined alphabet} associated with the system.

Consider a problem of coding of undetermined sequences such that the original undetermined sequence can be completely
reconstructed from the encoded sequence. One coding method refers to a \textit{binary representation}~\cite{sholomov13ran,sholomov13}
of undetermined alphabet, which is defined as a pair $(X, X^*)$ of $(N \times t)$-matrix $X$ with columns
$\x(i) \in \{0, 1\}^N, i \in [t],$ and $(N \times |\Tau|)$-matrix $X^*$ with columns $\x(T) \in \{0, 1, *\}^N, T \in \Tau,$ where
$\x(i)$ specifies $\x(T)$ in undetermined alphabet $\{0, 1, *\}$ if and only if~$i \in T$. Advantages of such method are
linear in $t$ complexity of the symbol reconstruction and the fact that the mentioned condition allows
to know only a small matrix $X$ for reconstruction of the original undetermined sequence while the matrix $X^*$ may contain
up to $2^t$ columns. Obviously, an SLD $s_1$-code $X = (\x(i), i \in [t])$ and the matrix $X^* = (\bigtri_{i \in T} \x(i), T \in \Tau)$
give the fairly compact binary representation of undetermined alphabet associated with the system
$\Tau = [t] \cup \{T \subset [t] | \; |T| \le s\}$~\cite{sholomov13}.

\subsection{Relations Between Parameters of LD $s_L$-Codes and SLD $s_L$-Codes}
\indent \indent
The following evident propositions from~\cite{dr81,d98,d03} associate the rate of LD $s_L$-codes~\req{RsL}
with the rate of SLD $s_L$-codes~\req{R*sL}.

\textbf{Proposition 1.}~\cite{dr81,d98,d03}.\quad
{\em Any LD $s_L$-code is the corresponding SLD $s_L$-code.}

\textbf{Proposition 2.}~\cite{dr81,d98,d03}.\quad
{\em Let $X = \| x_i(j) \|$ be an SLD $s_L$-code of length $N$ and size $t$.
Consider $(N \times t)$-matrix $X'=\|x'_i(j) \|$ with elements
$$
x_i'(j)\eq\begin{cases}
1, & \text{if $\;x_i(j)=0$},\cr
0, & \text{if $\;x_i(j)=1$}.\cr
\end{cases}
$$
Then the code of length $2N$ and size $t$
composed of all rows of the codes $X$ and $X'$ is an LD $s_L$-code.}

\textbf{Corollary 1.}~\cite{dr81,d98,d03}.\quad
{\em The rates of LD $s_L$-codes and SLD $s_L$-codes satisfy inequalities:}
\beq{LD-SLD}
R_L(s) \le R^*_L(s) \le 2 R_L(s).
\eeq

The next obvious proposition allows us to get another upper bound on the rate of SLD $s_L$-codes.

\textbf{Proposition 3.}\quad
{\em Let $X$ be an LD $s_L$-code of length $N$ and size $t$ with a codeword $\x(j_0)$ of weight $w$.
Then the code $X''$ of length $N-w$ and size $t-1$ constructed from the code $X$ by removing the codeword $\x(j_0)$ and
all rows $x_i$, for which $x_i(j_0) = 1$, is an LD $(s-1)_L$-code.}

\textbf{Corollary 2.}\quad
{\em The rate of SLD $s_L$-codes has the following upper bound:}
\beq{LD-SLD-new}
R^*_L(s) \le R_L(s-1).
\eeq

\textbf{Proof of Corollary 2.}\quad
Let $X$ be an arbitrary SLD $s_L$-code of length $N$ and size $t$. The code $X_1$ obtained in Proposition~2
from the code $X$ is a constant weight LD $s_L$-code of length $2N$, size $t$ and weight $N$. Then the code $X_2$
obtained in Proposition~3 from the code $X_1$ is an LD $(s-1)_L$-code of length $N$ and size $t-1$. Hence as $N \to \infty$
the inequality
$$
\frac{\log_2 [t-1]}{N} \le R_L(s-1) (1 + o(1))
$$
holds. It means correctness of~\req{LD-SLD-new}. $\square$

The best presently known lower and upper bounds on the rate $R_L(s)$ were recently obtained in~\cite{d14,d14_isit}.
The use of the inequalities~\req{LD-SLD} and~\req{LD-SLD-new}, the lower bound $\underline{R}_L(s)$~\cite{d14} and the upper
bound $\overline{R}_L(s)$~\cite{d14} on the rate of LD $s_L$-codes yields the results below.

\textbf{Theorem 2.}\quad {\large (}Relationship between $R^*_L(s)$ and $R_L(s)${\large )}
\newline
{\em The following three statements hold.}
\newline
{\em {\bf 1.} For any fixed $s \ge 2$ and $L \ge 1$ the rates $R^*_L(s)$ and $R_L(s)$ have relationship}
$$
R_L(s) \le R^*_L(s) \le \min \{ \; 2 R_L(s), \; R_L(s-1) \; \}.
$$
{\em {\bf 2.} For any fixed $L \ge 1$ and $s \to \infty$}
$$
R^*_L(s) = R_L(s) (1 + o(1)).
$$
{\em {\bf 3.} For any fixed $s \ge 2$ and $L \ge 1$ the rate of an SLD $s_L$-code satisfies the inequality}
%\beq{obviousR}
$$
\underline{R}_L(s) \le R^*_L(s) \le \overline{R}^*_L(s) \eq \min \{ \; 2 \overline{R}_L(s), \; \overline{R}_L(s-1) \; \}.
$$
%\eeq

\subsection{Random Coding Bounds on the Rate of SLD $s_L$-codes}
\indent \indent
In the given paper, we develop a random coding method based on the
ensemble of constant-weight codes and  establish  new lower random coding bounds on the rate of SLD $s_L$-codes.
Some of the methods which are used in the proof of the next theorem are presented in~\cite{d14,d14_isit}.

\textbf{Theorem 3.}\quad {\large (}Lower random coding bound $\underline{R}^*_L(s)${\large )}.\quad
\newline
{\em
The following three statements hold.
\newline
{\bf 1.} For any fixed $L \ge 1$ and $s \ge 2$ we have the inequality
\beq{lowerR}
R^*_L(s) \ge \underline{R}^*_L(s) \eq \max \limits_{0 < Q \le 1/2} \( h(Q) + \frac{B_L(s, Q)}{s + L - 1} \),
\eeq
where
\beq{lowerRFuncs}
\begin{split}
h(Q) &\eq -Q \log_2 Q - (1 - Q) \log_2 [1 - Q],\\
B_L(s, Q) &\eq Q \log_2 \[ \frac{p(1 - z)}{p(1 - z) + q(1 - z)}\] + (1 - Q) \log_2 \[ \frac{p(z)}{p(z) + q(z)} \],\\
p(z) &\eq z^s (z - z^s)^L,\\
q(z) &\eq (z - z^s) (1 - z^s - (1 - z)^s)^L,
\end{split}
\eeq
and $z$ is the unique root of the equation
\beq{QzEquation}
Q(p(z) + q(z)) = (1 - Q)(p(1 - z) + q(1 - z)).
\eeq
{\bf 2.} For fixed $L = 1, 2, \dots$ and $s \to \infty$
\beq{lowerRsAs}
\underline{R}^*_L(s) \ge \frac{L}{s^2 \log_2 e}(1 + o(1)).
\eeq
{\bf 3.} For fixed $s = 2, 3, \dots$ there exists a limit
\beq{lowerRLAs}
\underline{R}^*_{\infty}(s) \eq \lim_{L \to \infty} \underline{R}^*_L(s) = \log_2 \[ \frac{(s-1)^{s-1}}{s^s} + 1 \].
\eeq
If $s \to \infty$, then 
$$
\underline{R}^*_{\infty}(s) = \frac{\log_2 e}{es} (1 + o(1)) = \frac{0.5307 \dots}{s} (1 + o(1)).
$$
}

The numerical values of the lower bound~\req{lowerR}-\req{QzEquation} are shown in Table~\ref{tab1},
where the argument of maximum in~\req{lowerR} is denoted by $Q_L^*(s)$.
Note that the lower bound~\req{lowerR}-\req{QzEquation}  improves the random coding  bound obtained in~\cite{r89}
using the ensemble with independent binary symbols of codewords.
In addition one can see that for small values of $s \ge 2$ and $L \ge 1$, the lower bounds~\req{lowerR}-\req{QzEquation} are greater than
the lower bounds $\underline{R}_L(s)$ on the rate of LD $s_L$-codes from~\cite{d14}.

Note that, for $s \to \infty$, the asymptotic lower bound of $\underline{R}^*_L(s)$~\req{lowerRsAs} coincides
with the asymptotic behavior of the random coding bound on the rate of LD $s_L$-codes~\cite{d14}. In addition,
for $L \to \infty$, the asymptotics of $\underline{R}^*_L(s)$~\req{lowerRLAs} coincides
with the asymptotic behavior of the mentioned above bound from~\cite{d14}.

\begin{table}[ht]
\caption{Numerical values of the lower bound $\underline{R}^*_L(s)$}\label{tab1}
\begin{tabular*}{\textwidth}{@{\extracolsep{\fill}}ccccccc}
\hline
$s_L$ & $2_{1}$ & $2_{2}$ & $2_{3}$ & $2_{4}$ & $2_{5}$ & $2_{6}$ \cr
$\underline{R}_L^*(s)$ & $0.2075$ & $0.2457$ & $0.2635$ & $0.2744$ & $0.2819$ & $0.2874$ \cr
$Q_L^*(s)$ & $0.5000$ & $0.2764$ & $0.2432$ & $0.2297$ & $0.2228$ & $0.2180$ \cr
\hline
$s_L$ & $3_{1}$ & $3_{2}$ & $3_{3}$ & $3_{4}$ & $3_{5}$ & $3_{6}$ \cr
$\underline{R}_L^*(s)$ & $0.0800$ & $0.1153$ & $0.1348$ & $0.1470$ & $0.1552$ & $0.1611$ \cr
$Q_L^*(s)$ & $0.2000$ & $0.1794$ & $0.1686$ & $0.1613$ & $0.1561$ & $0.1524$ \cr
\hline
$s_L$ & $4_{1}$ & $4_{2}$ & $4_{3}$ & $4_{4}$ & $4_{5}$ & $4_{6}$ \cr
$\underline{R}_L^*(s)$ & $0.0439$ & $0.0684$ & $0.0838$ & $0.0941$ & $0.1014$ & $0.1068$ \cr
$Q_L^*(s)$ & $0.1479$ & $0.1391$ & $0.1326$ & $0.1275$ & $0.1234$ & $0.1201$ \cr
\hline
$s_L$ & $5_{1}$ & $5_{2}$ & $5_{3}$ & $5_{4}$ & $5_{5}$ & $5_{6}$ \cr
$\underline{R}_L^*(s)$ & $0.0279$ & $0.0456$ & $0.0575$ & $0.0660$ & $0.0723$ & $0.0771$ \cr
$Q_L^*(s)$ & $0.1209$ & $0.1150$ & $0.1103$ & $0.1064$ & $0.1030$ & $0.1003$ \cr
\hline
$s_L$ & $6_{1}$ & $6_{2}$ & $6_{3}$ & $6_{4}$ & $6_{5}$ & $6_{6}$ \cr
$\underline{R}_L^*(s)$ & $0.0194$ & $0.0325$ & $0.0420$ & $0.0490$ & $0.0544$ & $0.0587$ \cr
$Q_L^*(s)$ & $0.1027$ & $0.0983$ & $0.0947$ & $0.0915$ & $0.0889$ & $0.0865$ \cr
\hline
\end{tabular*}
\end{table}

\section{Proof of Theorem 3}
\indent \indent
This Section contains five lemmas that are only stated.
The proofs of Lemma~1-5 are presented in Appendix.

\textbf{Proof of Statement 1.}\quad
Fix $L \ge 1$, $s \ge 2$ and a parameter $Q, \; 0 < Q \le 1/2$. The bound~\req{lowerR}-\req{QzEquation} is obtained by the method of random coding over the ensemble of binary constant-weight codes~\cite{drr89} defined as the ensemble $E(N, t, Q)$ of binary codes $X$ of length $N$ and size $t$, where the codewords are chosen independently and equiprobably from the set consisting of all ${N \choose \lfloor QN \rfloor}$ codewords of a fixed weight $\lfloor QN \rfloor$.
A pair of sets $(\S, \L), |\S| = s, |\L| = L, \S \cap \L = \varnothing,$ we call an {\em $(s_L^*)$-bad} pair if
$$
\bigtri_{i \in \S} \x(i) \trianglerighteq \bigtri_{j \in \L} \x(j).
$$
For the ensemble $E(N, t, Q)$, denote by $P(N, Q, s, L)$ the probability of the event ``the pair $(\S, \L)$ is $(s_L^*)$-bad''. Note that the absence of $(s_L^*)$-bad pair of subsets in the code is the criterion of SLD $s_L$-code. Hence, similarly to the arguments in the proof of the lower random coding bound on the rate $R_L(s)$~\req{RsL} in~\cite{d14}, the rate $R^*_L(s)$~\req{R*sL} satisfies the inequality
\beq{lowerRbase}
\begin{split}
R^*_L(s) \ge \underline{R}^*_L(s) \eq \frac{1}{s + L - 1} \max_{0 < Q < 1} A^*_L(s, Q),\\
A^*_L(s, Q) \eq \varlimsup_{N \to \infty} \frac{- \log_2 P(N, Q, s, L)}{N}.
\end{split}
\eeq

Note that the set of all $s_L^*$-bad pairs of any codeword weight is invariant under the binary negation operation, it implies the equality $P(N, Q, s, L) = P(N, 1-Q, s, L)$. Therefore, it is enough to consider only $0 < Q \le 1/2$.

To complete the proof of the theorem, it is sufficient to compute the function $A^*_L(s, Q)$~\req{lowerRbase}.

\textbf{Lemma 1.}\quad
{\em If there exists a solution $z, 0 < z < 1,$ of the equation~\req{QzEquation}, then the function $A^*_L(s, Q)$~\req{lowerRbase} equals}
\beq{Avalue}
(s + L - 1) h(Q) + (1 - Q) \log_2 \[ \frac{p(z)}{p(z) + q(z)} \] + Q \log_2 \[ \frac{p(1 - z)}{p(1 - z) + q(1 - z)} \],
\eeq
{\em where the functions $h(\cdot), p(\cdot)$ and $q(\cdot)$ are determined by \req{lowerRFuncs}.}

\textbf{Lemma 2.}\quad
{\em The function}
\beq{rhoz}
\rho(z) \eq \frac{p(z) + q(z)}{p(1-z) + q(1-z)}, \quad 0 < z < 1,
\eeq
{\em continuously maps the interval $(0, 1)$ into the interval $(0, +\infty)$ and strictly increases.}

By Lemma~2 the equation~\req{QzEquation} has the unique solution. Thus,
the condition of Lemma~1 is clear, it means that the bound~\req{lowerR}-\req{QzEquation} is proved. $\quad \square$

\textbf{Proof of Statement 2.}\quad
For fixed $s \ge 2$ and $L \ge 1$, let us interpret equation~\req{QzEquation} as a function $Q_L(s, z)$ of the argument $z, 0 < z < 1$, i.e.,
\beq{QOfz}
Q_L(s, z) \eq \frac{p(1-z) + q(1-z)}{p(1-z) + q(1-z) + p(z) + q(z)},
\eeq
where the functions $p(\cdot)$ and $q(\cdot)$ are determined in~\req{lowerRFuncs}.

Due to existence and uniqueness of the root of the equation~\req{QzEquation}, continuity and monotonicity of the function~\req{QOfz} (by Lemma~2),
one can rewrite the definition of the random coding bound \req{lowerR}-\req{QzEquation} as
\beq{lowerRByz}
\underline{R}^*_L(s) \eq \max \limits_{1 / 2 \le z < 1} T_L(s, z),
\eeq
where
\beq{TOfz}
T_L(s, z) \eq h(Q_L(s, z)) + B_L(s, Q_L(s, z)).
\eeq

Let $L \ge 1$ be fixed and $s \to \infty$. If in definition~\req{TOfz} we put $z = 1 - \lambda / s$, where the parameter $\lambda = \lambda_L$ is independent of $s$, then \req{lowerRByz}~means that
\beq{sIneq}
\underline{R}^*_L(s) \ge T_L \( s, 1 - \frac{\lambda}{s} \) .
\eeq

\textbf{Lemma 3.}\quad
{\em For a fixed $L \ge 1$ and $s \to \infty$, the next asymptotic equality holds:}
\beq{TAs}
T_L \( s, 1 - \frac{\lambda}{s} \) = \frac{L}{s^2} \( - \lambda \log_2 [ 1 - e^{-\lambda} ]\) (1 + o(1)).
\eeq

Taking derivative one can check that at $\lambda = \frac{1}{\log_2 e}$ the maximum
\beq{max}
\max \limits_{\lambda > 0} \left\{ - \lambda \log_2 [ 1 - e^{-\lambda} ] \right\} = \frac{1}{\log_2 e}
\eeq
is attained. Therefore, \req{sIneq} and \req{max} imply for the random coding bound~\req{lowerR}-\req{QzEquation} the asymptotic inequality~\req{lowerRsAs}. $\quad \square$

\textbf{Proof of Statement 3.}\quad
For fixed $s \ge 2$ and $L \ge 1$, let us introduce the following function
\beq{gOfz}
g(z) \eq g_L(s, z) = \frac{z - z^s}{1 - z - (1 - z)^s}, \quad \frac{1}{2} \le z < 1.
\eeq
It is clear that $g(z)$~\req{gOfz} monotonically increases in the interval $[ 1 / 2, 1 )$, attains $1$ at the point $z = \frac{1}{2}$ and has the left limit $s - 1$ as $z \to 1$.

For large enough parameter $L$ and a fixed parameter $c > 0$ independent of $L$, one can see that the root of equation
\beq{zEquation}
\( \frac{g(z)}{1 + g(z)} \)^L = c (1-z), \quad \frac{1}{2} \le z < 1,
\eeq
exists and is unique, since the left-hand side of~\req{zEquation} monotonically increases and the right-hand side of~\req{zEquation} strictly decreases. Denote this root by $z_L(s, c)$.

Let $s \ge 2$ be fixed and $L \to \infty$.

\textbf{Lemma 4.}\quad
{\em The substitution of $z = z_L(s, c)$ into the function~\req{TOfz} yields}
\begin{multline}
\label{TAsL}
T_L(s, z_L(s, c)) \cdot (1 + o(1)) = \log_2 [s + c] - \frac{s + c - 1}{s + c} \log_2 [s + c - 1] +\\
+ \frac{1}{s + c} \log_2 \[ \frac{(s-1)^{s-1}}{s^s} \], \quad
L \to \infty.
\end{multline}

The definition~\req{lowerRByz} means that
\beq{LIneq}
\underline{R}^*_L(s) \ge T_L(s, z_L(s, c)) (1 + o(1)), \quad L \to \infty, \quad \forall \, c = c(s) > 0.
\eeq

Calculating the derivative in $c$, one can check that maximum of the right-hand side of~\req{TAsL} is attained at the point $c = c(s) = \frac{s^s - (s-1)^s}{(s-1)^{s-1}}$. If we substitute this value $c = c(s)$ into~\req{TAsL}, then the use of~\req{LIneq} establishes for the random coding bound~\req{lowerR}-\req{QzEquation} the inequality
\beq{lowerRLAsIneq}
\underline{R}^*_L(s) \ge \log_2 \[ \frac{(s-1)^{s-1}}{s^s} + 1 \] (1 + o(1)), \quad L \to \infty.
\eeq

\textbf{Lemma 5.}\quad
{\em The asymptotic inequality~\req{lowerRLAsIneq} is an equality.}

Statement~3 of Theorem~3 is proved. $\quad \square$

\appendix
\section{Proofs of Lemma~1-5}

\indent \indent
\textbf{Proof of Lemma 1.}\quad
Let us use the terminology of {\em types}~\cite{ck81}. Consider an arbitrary set of size $s$ consisting of binary codewords of length $N$ and weight $\lfloor QN \rfloor$:
$(\x(1), ..., \x(s))$, where $\x(i) \in \{0, 1\}^N, ~\forall i \in [s]$.
The set forms $(N \times s)$-matrix $X_s$. Let $\a \eq (a_1, ..., a_s) \in \{0, 1\}^s$. Denote a {\em type} of the matrix $X_s$ by $\{n(\a)\}$, where $n(\a), 0 \le n(\a) \le N$ is the number of $\a$-rows in the matrix $X_s$. Obviously, for any matrix $X_s$ we have
$$
\sum_{\a} n(\a) = N.
$$
By $n(\0)$ ($n(\1)$) denote the number of the rows in $X_s$ consisting of all zeros (ones). It allows to represent $P(N, Q, s, L)$ as
\beq{PsLbad}
P(N, Q, s, L) = \sum_{\{n(\a)\} \in \N} \frac{N!}{\prod_{\a} n(\a)!} {N - n(\0) - n(\1) \choose \lfloor QN \rfloor - n(\1)}^L {N \choose \lfloor QN \rfloor}^{-s-L},
\eeq
where the set $\N$ consists of all possible types ${n(\a)}, \a \in \{0, 1\}^s$, such that:
\beq{sumSet}
\begin{gathered}
0 \le n(\a) \le N \quad \forall \, \a \in \{0, 1\}^s, \quad
n(\0) \le N - \lfloor QN \rfloor, \quad
n(\1) \le \lfloor QN \rfloor,\\
\sum_{\a} n(\a) = N, \quad
\sum_{\a : \; a_i = 1} n(\a) = \lfloor QN \rfloor \quad \forall \, i \in [s].
\end{gathered}
\eeq

Let $N \to \infty$. For every type ${n(\a)}, \, \a \in \{0, 1\}^s$, let us consider the corresponding distribution $\tau \eq \{\tau(\a)\} : \tau(\a) = \frac{n(\a)}{N}$.
Thus, for $N \to \infty$, the set $\N$ accords with the set $\Tau$ consisting of the distributions with the following properties induced by~\req{sumSet}:
\beq{minSet}
\tau \in \Tau \Longleftrightarrow
\left\{ \begin{gathered}
0 \le \tau(\a) \le 1 \quad \forall \, \a \in \{0, 1\}^s, \quad
\tau(\0) \le 1 - Q, \quad
\tau(\1) \le Q,\\
\sum_{\a \in \{0, 1\}^s} \tau(\a) = 1, \quad
\sum_{\a : \; a_i = 1} \tau(\a) = Q \quad \forall \, i \in [s].
\end{gathered} \right\}
\eeq

Applying the Stirling approximation, we obtain the following logarithmic asymptotic behavior of the summand in the sum~\req{PsLbad} for $\tau \in \Tau$:
\begin{align}
\notag
\begin{split}
-\log_2 \sum_{\{n(\a)\} \in \N} \frac{N!}{\prod_{\a} n(\a)!} {N - n(\0) - n(\1) \choose \lfloor QN \rfloor - n(\1)}^L {N \choose \lfloor QN \rfloor}^{-s-L} =\\
= N F(\tau, Q) (1 + o(1)), \quad \text{where},
\end{split}
\\
\label{FtQ}
\begin{split}
F(\tau, Q) \eq \sum_{\a} \tau(\a) \log_2 [\tau(\a)] - (1 - \tau(\0) - \tau(\1)) L h\( \frac{Q - \tau(\1)}{1 - \tau(\0) - \tau(\1)} \) +\\
+ (s + L) h(Q).
\end{split}
\end{align}

For the given $Q$, let the minimum of the function $F(\tau, Q)$ be attained at $\tau_Q = \{\tau_Q(\a)\}$, then
\beq{AAsMin}
A^*_L(s, Q) \eq \varlimsup_{N \to \infty} \frac{-\log_2 P(s, L, Q, N)}{N} = F(\tau_Q, Q) = \min_{\tau \in \Tau} F(\tau, Q).
\eeq

Since $F$ is continuous in the admissible compact space $\Tau$, finding the minimum of $F$
under constraints~\req{minSet} with excluded boundaries is sufficient to calculate \req{AAsMin}. Let us write the minimization problem: $F \to \min$,
\begin{align}
\label{SearchDomain}
\text{Search domain $\mathbb{T}$:} \hspace{1.5cm} &0 < \tau(\a) < 1 \quad \forall \, \a \in \{0, 1\}^s, \quad \tau(\1) < Q, \quad \tau(\0) < 1 - Q,\\
\label{Restrictions}
\text{Restrictions:} \hspace{1.5cm}
&\begin{cases}
&\displaystyle{\sum_{\a \in \{0, 1\}^s} \tau(\a) = 1,}\\
&\displaystyle{\sum_{\a : \; a_i = 1} \tau(\a) = Q \quad \forall \, i \in [s],}
\end{cases}\\
\label{MainFunction}
\text{Main Function:} \hspace{1.5cm} &F(\tau, Q) = \req{FtQ} : \; \mathbb{T} \to \mathbb{R}.
\end{align}

To find the extremal distribution $\tau_Q$ we apply the standard Lagrange multipliers method. Consider the Lagrangian:
\beq{Lagrangian}
\Lambda \eq F(\tau, Q) + \lambda_0 \( \sum_{\a \in \{0, 1\}^s} \tau(\a) - 1 \) + \sum_{i = 1}^{s} \lambda_i \( \sum_{\a: a_i = 1} \tau(\a) - Q \).
\eeq
The necessary conditions for the extremal distribution $\tau_Q$ are:
\beq{NecCondLong}
\begin{cases}
\frac{\partial \Lambda}{\partial(\tau(\a))} &= \log_2 [\tau(a)] + \log_2 e + \lambda_0 + \sum_{i = 1}^{s} \lambda_i a_i = 0, \quad \forall \; \a \in \{0, 1\}^s \setminus \{\0, \1\},\\
\frac{\partial \Lambda}{\partial(\tau(\0))} &= \log_2 [\tau(\0)] + \log_2 e + \lambda_0 + L \log_2 \[ \frac{1 - \tau(\0) - \tau(\1)}{1 - Q - \tau(0)} \] = 0,\\
\frac{\partial \Lambda}{\partial(\tau(\1))} &= \log_2 [\tau(\1)] + \log_2 e + \lambda_0 + \sum_{i = 1}^{s} \lambda_i + L \log_2 \[ \frac{1 - \tau(\0) - \tau(\1)}{Q - \tau(\1)} \] = 0.
\end{cases}
\eeq

Let us show that the matrix of second derivatives of the Lagrangian is positive definite. Indeed, we have
\begin{align*}
\frac{\partial^2 \Lambda}{\partial(\tau(\a))^2} &= \frac{\log_2 e}{\tau(\a)} > 0, \quad \forall \; \a \in \{0, 1\}^s \setminus \{\0, \1\},\\
\frac{\partial^2 \Lambda}{\partial(\tau(\0))^2} &= \frac{\log_2 e}{\tau(\0)} + L \log_2 e \frac{Q - \tau(\1)}{(1 - \tau(\0) - \tau(\1))(1 - Q - \tau(\0))} > 0,\\
\frac{\partial^2 \Lambda}{\partial(\tau(\1))^2} &= \frac{\log_2 e}{\tau(\1)} + L \log_2 e \frac{1 - Q - \tau(\0)}{(1 - \tau(\0) - \tau(\1))(Q - \tau(\1))} > 0,\\
\frac{\partial^2 \Lambda}{\partial(\tau(\0))\partial(\tau(\1))} &= -L \log_2 e \frac{1}{1 - \tau(\0) - \tau(\1)} < 0,
\end{align*}
and the other elements of the matrix are zeros. That is why, this matrix is positive definite. Note that the matrix of second derivatives of the function $F(\tau, Q)$ coincides with the above matrix. Therefore~\cite{opu}, $F$ is strictly $\cup$-convex in the domain $\mathbb{T}$. Moreover, the constraint equations~\req{Restrictions} define an affine subspace $\mathbb{G}$ in $\mathbb{R}^{2^s}$ of dimension $(2^s - s - 1)$, that is why $F$ is strictly $\cup$-convex in $\mathbb{T} \cap \mathbb{G}$. Hence a local minimum of $F$ in $\mathbb{T} \cap \mathbb{G}$ is global and unique. Due to the Karush-Kuhn-Tacker theorem~\cite{opu}, it is clear that each solution satisfying the system~\req{NecCondLong} and the constraints~\req{Restrictions} is unique and gives the desired minimum distribution $\tau_Q$ for $F(\tau, Q)$.

Note that the symmetry of the problem yields equality: $\nu \eq \lambda_1 = \lambda_2 = ... = \lambda_s$. To prove this, we need to check that $\lambda_i = \lambda_j$ for $i \not= j$. Let $\bar\a_i \eq (0, \dots, 1, \dots, 0)$ be a row of length $s$, which has $1$ at the $i$-th position and $0's$ at the other positions. A permutation of indices $i$ and $j$ leads to an equivalent problem. Hence, if $\tau_Q^1$ is a solution, then $\tau_Q^2$ is also a solution, where $\tau_Q^2(\a) \eq \tau_Q^1(\tilde \a)$ and $\tilde \a$ is a row, obtained by permutation of indices $i$ and $j$ from the row $\a$. The uniqueness of the solution $\tau_Q$ implies that the distribution $\tau_Q^1$ coincides with the distribution $\tau_Q^2$. In particular, $\tau_Q^1(\bar \a_i) = \tau_Q^2(\bar \a_i) = \tau_Q^1(\bar \a_j)$. From the first equation of~\req{NecCondLong}, it follows
that $\lambda_i = \lambda_j$.

Introduce a parameter $\mu \eq e 2^{\lambda_0}$. Then the equations~\req{NecCondLong} have the form:
\beq{NecCond}
\begin{cases}
\log_2 \mu + \log_2 [\tau(\a)] + \nu \sum_{i = 1}^{s} a_i = 0,
\vspace{0.1cm}\\
\log_2 \mu + \log_2 [\tau(\0)] + L \log_2 \[ \frac{1 - \tau(\0) - \tau(\1)}{1 - Q - \tau(\0)} \] = 0,
\vspace{0.1cm}\\
\log_2 \mu + \log_2 [\tau(\1)] + L \log_2 \[ \frac{1 - \tau(\0) - \tau(\1)}{Q - \tau(\1)} \] + s \nu = 0.
\end{cases}
\eeq

After substitution $z \eq \frac{1}{1 + 2^{-\nu}}, 0 < z < 1$, the first equation of~\req{NecCond} gives
\beq{tau_a}
\tau(\a) = \frac{2^{-\nu \sum a_i}}{\mu} = \frac{1}{\mu z^s} (1 - z)^{\sum a_i} z^{s - \sum a_i} \quad \forall \; \a \in \{0, 1\}^s \setminus \{\0, \1\}.
\eeq

Substitution~\req{tau_a} into the first and the second equations of the system~\req{Restrictions} leads to
\begin{align}
\label{Rest01}
1 &= \frac{1}{\mu z^s} \sum_{i = 1}^{s-1} {s \choose i} z^i (1 - z)^{s - i} + \tau(\0) + \tau(\1) = \frac{1 - z^s - (1 - z)^s}{\mu z^s} + \tau(\0) + \tau(\1),\\
\label{Rest1}
Q &= \frac{1}{\mu z^s} \sum_{i = 1}^{s-1} {s - 1 \choose i} z^i (1 - z)^{s - i} + \tau(\1) = \frac{1 - z - (1 - z)^s}{\mu z^s} + \tau(\1),
\end{align}
correspondingly. Subtraction~\req{Rest1} from~\req{Rest01} yields
\beq{Rest0}
1 - Q = \frac{z - z^s}{\mu z^s} + \tau(\0).
\eeq

Due to~\req{Rest01}-\req{Rest0} the second and third equations of the system~\req{NecCond} are equivalent to
\beq{NecCondt01}
\begin{split}
\mu \( 1 - Q - \frac{z - z^s}{\mu z^s} \) \( \frac{1 - z^s - (1 - z)^s}{z - z^s} \)^L &= 1,\\
\mu \( Q - \frac{1 - z - (1 - z)^s}{\mu z^s} \) \( \frac{1 - z^s - (1 - z)^s}{1 - z - (1 - z)^s} \)^L &= 1,
\end{split}
\eeq
respectively.

To shorten the formulas let us introduce the functions of the parameters $s$, $L$ and $z$:
\beq{Shorten}
\begin{split}
p(z) &\eq p_L(s, z) = z^s \(z - z^s\)^L,\\
q(z) &\eq q_L(s, z) = (z - z^s) (1 - z^s - (1 - z)^s)^L,\\
r(z) &\eq r_L(s, z) = z^s (1 - z^s - (1 - z)^s)^L.
\end{split}
\eeq

The use of such notations yields the following expressions of $\mu$ from the both equations~\req{NecCondt01}:
\begin{align}
\label{MuThrough0}
\mu &= \frac{1}{1 - Q} \frac{p(z) + q(z)}{r(z)},\\
\label{MuThrough1}
\mu &= \frac{1}{Q} \frac{p(1 - z) + q(1 - z)}{r(z)}.
\end{align}

Equating of~\req{MuThrough0} and~\req{MuThrough1} leads to the equation on the parameter $z$:
$$
Q (p(z) + q(z)) = (1 - Q) (p(1 - z) + q(1 - z)),
$$
which coincides with the equation~\req{QzEquation}.
%\{it's obvious that there exists the root; uniqueness isn't obvious\}

The substitutions~\req{MuThrough0} into~\req{Rest0} and~\req{MuThrough1} into~\req{Rest1} give:
\beq{tau_01}
\begin{split}
\tau(\0) &= (1 - Q) \frac{p(z)}{p(z) + q(z)},\\
\tau(\1) &= Q \frac{p(1 - z)}{p(1 - z) + q(1 - z)}.
\end{split}
\eeq

So, let us calculate the value of $F(\tau, Q)$~\req{FtQ}, where the distribution $\tau$ is specified by~\req{tau_a} and~\req{tau_01}. At the beginning, we compute the following sum:
\begin{align}
\notag
&\sum_{\a: \; \a \neq \0, \1} \tau(\a) \log_2 [\tau(\a)] = \{ \text{by~\req{tau_a}} \} =\\
\notag
&= \sum_{i = 1}^{s - 1} {s \choose i} \frac{1}{\mu z^s} (1 - z)^{s - i} z^i \( \log_2 \[ \frac{1}{\mu z^s} \] + i \log_2 z + (s - i) \log_2 [1 - z] \) =\\
\notag
&= \frac{1 - z^s - (1 - z)^s}{\mu z^s} \log_2 \[ \frac{1}{\mu z^s} \] + \frac{z - z^s}{\mu z^s} \log_2 \[ z^s \] + \frac{1 - z - (1 - z)^s}{\mu z^s} \log_2 \[ (1 - z)^s \] =\\
\notag
&= \{ \text{by~\req{Rest01},~\req{Rest0} and~\req{Rest1}} \} =\\
\notag
&= (1 - \tau(\0) - \tau(\1)) \log_2 \[ \frac{1}{\mu z^s} \] + (1 - Q - \tau(\0)) \log_2 \[ z^s \] + (Q - \tau(\1)) \log_2 \[ (1 - z)^s \] =\\
\label{sumA}
&= (1 - Q - \tau(\0)) \log_2 \[ \frac{1}{\mu} \] + (Q - \tau(\1)) \log_2 \[ \frac{(1 - z)^s}{\mu z^s} \].
\end{align}

Further, the use of~\req{sumA} implies
\begin{align}
\notag
\sum_{\a: \; \a \neq \0, \1} &\tau(\a) \log_2 [\tau(\a)] - (1 - \tau(\0) - \tau(\1)) L h\( \frac{Q - \tau(\1)}{1 - \tau(\0) \tau(\1)} \) =\\
\notag
&= (1 - Q - \tau(\0)) \( - \log_2 \mu - L \log_2 \[ \frac{1 - \tau(\0) - \tau(\1)}{1 - Q - \tau(\0)} \] \) +\\
\notag
&+ (Q - \tau(\1)) \( -\log_2 \mu - \log_2 \[ \frac{z^s}{(1 - z)^s} \] - L \log_2 \[ \frac{1 - \tau(\0) - \tau(\1)}{Q - \tau(\1)} \] \) =\\
\notag
&= \{ \text{by~\req{NecCond}}\} =\\
\label{sumB}
&= (1 - Q - \tau(\0)) \log_2 [\tau(\0)] + (Q - \tau(\1)) \log_2 \tau(\1).
\end{align}

Finally, the use of~\req{sumB} and~\req{tau_01} leads to
\begin{align}
\notag
F(\tau, Q) &= (s + L) h(Q) + (1 - Q) \log_2 [\tau(\0)] + Q \log_2 [\tau(\1)] = \req{Avalue}.
\end{align}
Thus, Lemma~1 is proved. $\quad \square$

\textbf{Proof of Lemma 2.}\quad
Let us rewrite the formula~\req{rhoz} using the monotonically increasing function $g(z)$~\req{gOfz}:
\beq{rhoz2}
\rho(z) = \frac{z^s (g(z))^L + (z - z^s) (1 + g(z))^L}{(1-z)^s + (1 - z - (1-z)^s) (1 + g(z))^L}.
\eeq
The devision of the numerator and the denominator of~\req{rhoz2} by $(z - z^s)(1 + g(z))^L$ leads to
$$
\rho(z) = \frac{ \( \frac{g(z)}{1 + g(z)} \)^L \cdot \frac{z^s}{z - z^s} + 1}{\frac{(1-z)^s}{z - z^s} \cdot \frac{1}{(1+g(z))^L} + \frac{1}{g(z)}},
$$
where the function $\frac{z^s}{z - z^s}$ is strictly increasing and the function $\frac{(1-z)^s}{z - z^s}$ is strictly decreasing.
Thus, it is clear that $\rho(z)$ is strictly increasing.

Note that $g(z) \to \frac{1}{s-1}$ as $z \to 0$ and $g(z) \to s-1$ as $z \to 1$. Therefore, by~\req{rhoz2} the following limits are true:
\begin{align*}
&\lim_{z \to 0 + 0} \rho(z) = 0,\\
&\lim_{z \to 1 - 0} \rho(z) = +\infty.
\end{align*}
Lemma~2 is proved. $\quad \square$

\textbf{Proof of Lemma 3.}\quad
Let us introduce the following notations:
\beq{UV}
\begin{split}
U_L(s, z) &\eq \frac{p(1-z)}{p(1-z) + q(1-z)},\\
V_L(s, z) &\eq \frac{p(z)}{p(z) + q(z)}.
\end{split}
\eeq
Then the function~\req{TOfz} can be represented as
\beq{TOfz2}
T_L(s, z) = - Q \log_2 Q - (1 - Q) \log_2 [1 - Q] + \frac{1}{s + L - 1} \( Q \log_2 U + (1 - Q) \log_2 V \),
\eeq
where the shorthands $Q = Q_L(s, z)$, $U = U_L(s, z)$ and $V = V_L(s, z)$ are used.

Computation of two first terms of asymptotic expansions of $p(z), q(z),$ $p(1-z), q(1-z)$~\req{lowerRFuncs} for $z = 1 - \lambda / s$ and $s \to \infty$ leads to the equalities
\beq{pqAs}
\begin{split}
p(1 - z) = p \( \frac{\lambda}{s} \) &= \( \frac{\lambda}{s} \)^{s+L} - \( \frac{\lambda}{s} \)^{s(L+1)},\\
q(1 - z) = q \( \frac{\lambda}{s} \) &= \frac{\lambda (1 - e^{-\lambda})^L}{s} + \frac{L \lambda^3 e^{-\lambda} (1 - e^{-\lambda})^{L-1}}{2s^2} + o\( \frac{1}{s^2} \),\\
p(z) = p \( 1 - \frac{\lambda}{s} \) &= e^{\lambda} (1 - e^{-\lambda})^L + \frac{\lambda e^{-\lambda} (1 - e^{-\lambda})^L (\lambda + L \lambda - 2L e^\lambda - \lambda e^\lambda)}{2 (e^\lambda - 1) s} + o\( \frac{1}{s} \),\\
q(z) = q \( 1 - \frac{\lambda}{s} \) &= (1 - e^{-\lambda})^L + \frac{\lambda e^{-\lambda} (1 - e^{-\lambda})^L (\lambda + L \lambda - 2 e^\lambda)}{2s} + o\( \frac{1}{s} \).
\end{split}
\eeq

Using~\req{pqAs}, one can obtain the following asymptotic equalities for the expressions~\req{QOfz},\req{UV}
\beq{QUVAs}
\begin{split}
Q_L \( s, 1 - \frac{\lambda}{s} \) &= \frac{\lambda}{s} + \frac{L \lambda^2}{(e^\lambda - 1) s^2} + o\( \frac{1}{s^2} \),\\
U_L \( s, 1 - \frac{\lambda}{s} \) &= \( \frac{\lambda}{s} \)^{s + L - 1} (1 - e^{-\lambda})^{-L} \( 1 + o\( \frac{1}{s} \) \),\\
V_L \( s, 1 - \frac{\lambda}{s} \) &= e^{-\lambda} \( 1 + \frac{\lambda - L \lambda - \lambda^2 / 2}{s} + o\( \frac{1}{s} \) \).
\end{split}
\eeq

Finally, equalities~\req{QUVAs} yield the asymptotic behavior of~\req{TOfz2} that coincides with~\req{TAs}. $\quad \square$

\textbf{Proof of Lemma 4.}\quad
Let $s \ge 2$ be fixed and $L \to \infty$. It is obvious that
\beq{zgAs}
\begin{split}
z_L(s, c) &= 1 + o(1), \quad \text{and hence},\\
g(z_L(s, c)) &= (s - 1)(1 + o(1)).
\end{split}
\eeq

The use of definitions~\req{lowerRFuncs} and division of upper and lower parts of fractions~\req{QOfz},\req{UV} by $(1 - z - (1 - z)^s)$ allow us to rewrite expressions $Q$, $U$ and $V$~\req{QOfz},\req{UV} in a more convenient form
\beq{QUVOfg}
\begin{split}
Q_L(s, z) &= \frac{(1-z)^s + (1 - z - (1-z)^s) (1 + g(z))^L}{(1 - z)^s + (1 - z^s - (1 - z)^s) (1 + g(z))^L + z^s (g(z))^L},\\
U_L(s, z) &= \frac{(1-z)^s}{(1-z)^s + (1 - z - (1-z)^s) (1 + g(z))^L},\\
V_L(s, z) &= \frac{z^s (g(z))^L}{z^s (g(z))^L (z - z^s) (1 + g(z))^L}.
\end{split}
\eeq

The equalities~\req{zgAs}-\req{QUVOfg} imply the following asymptotics
\beq{QUVAsL}
\begin{split}
Q_L(s, z_L(s, c)) &= \frac{1}{s+c} (1 + o(1)),\\
U_L(s, z_L(s, c)) &= \( \frac{(s-1)^{s-1}}{s^s} \)^L (1 + o(1)),\\
V_L(s, z_L(s, c)) &= \frac{1}{1 + \frac{s}{c}} (1 + o(1)).
\end{split}
\eeq

Next, the substitution~\req{QUVAsL} into the expression~\req{TOfz2} involves~\req{TAsL}. $\quad \square$

\textbf{Proof of Lemma 5.}\quad
To prove the equality sign in~\req{lowerRLAsIneq}, let us denote arbitrary sequence of argument of maximum~\req{lowerRByz} by $z = z_L(s), \; 1/2 \le z_L(s) < 1$.
We will consider some cases and find a contradictions with~\req{lowerRLAsIneq}.
First, suggest that the sequence $z_L(s)$ is bounded by a constant $d < 1$, i.e., $1/2 \le z_L(s) \le d < 1$. Then due to~\req{QUVOfg} the asymptotic equalities
\beq{caseBounded}
\begin{split}
Q_L(s, z_L(s)) &= \frac{1}{1 + g(z)} (1 + o(1)),\\
U_L(s, z_L(s)) &= \frac{(1 - z)^s}{1 - z - (1 - z)^s} \frac{1}{(1 + g(z))^L} (1 + o(1)),\\
V_L(s, z_L(s)) &= \frac{z^s}{z - z^s} \( \frac{g(z)}{1 + g(z)} \)^L (1 + o(1)), \quad L \to \infty,
\end{split}
\eeq
hold. However, the computation of asymptotic behavior of $T_L(s, z_L(s))$~\req{TOfz2}, using~\req{caseBounded}, yields $\underline{R}^*_L(s) = T_L(s, z_L(s)) \to 0$ as $L \to \infty$. The current case involves the contradiction with~\req{lowerRLAsIneq}. Hence, it is clear without less of generality that $z_L(s) \to 1$ (\req{zgAs} holds).

Further, let us assume that
\beq{assumeToZero}
\( \frac{g(z)}{1 + g(z)} \)^L \frac{1}{1 - z} \to 0, \quad L \to \infty.
\eeq
Then using~\req{zgAs} and~\req{assumeToZero} one can achive the following asymptotic behaviors of~\req{QUVOfg}
\beq{caseToZero}
\begin{split}
Q_L(s, z_L(s)) &= \frac{1}{s} (1 + o(1)),\\
U_L(s, z_L(s)) &= \frac{(1 - z)^{s-1}}{(1 + g(z))^L} (1 + o(1)),\\
V_L(s, z_L(s)) &= \frac{1}{s} \( \frac{g(z)}{1 + g(z)} \)^L \frac{1}{1 - z} (1 + o(1)), \quad L \to \infty.
\end{split}
\eeq
Nevertheless, the equalities~\req{zgAs}~and~\req{caseToZero} leads to $\underline{R}^*_L(s) = T_L(s, z_L(s)) \to 0$ as $L \to \infty$. So, the current case has the contradiction with~\req{lowerRLAsIneq}.

Next, let us assume that
\beq{assumeToInfty}
\( \frac{g(z)}{1 + g(z)} \)^L \frac{1}{1 - z} \to \infty, \quad L \to \infty.
\eeq
The use of~\req{zgAs} and~\req{assumeToInfty} leads to the following asymptotic behavior of~\req{QUVOfg}
\beq{caseToInfty}
\begin{split}
Q_L(s, z_L(s)) &= \( \frac{1 + g(z)}{g(z)} \)^L (1 - z) (1 + o(1)),\\
U_L(s, z_L(s)) &= \frac{(1 - z)^{s-1}}{(1 + (z))^L} (1 + o(1)),\\
V_L(s, z_L(s)) &= 1 + o(1), \quad L \to \infty.
\end{split}
\eeq
It is obvious that the equalities~\req{zgAs}~and~\req{caseToInfty} yield
\beq{complex}
T_L(s, z_L(s)) = \frac{Q(s-1)}{s + L - 1} \log_2 [1 - z] + o(1).
\eeq
One can see that from the first equality in~\req{caseToInfty} it follows that
$$
Q = O(1 - z).
$$
Therefore, the asymptotic equality~\req{complex} implies $\underline{R}^*_L(s) = T_L(s, z_L(s)) \to 0$ as $L \to \infty$. Therefore, the current case has the contradiction with~\req{lowerRLAsIneq}.

Without loss of generality we can conclude that
\beq{zEquationAs}
\( \frac{g(z)}{1 + g(z)} \)^L = c (1-z) (1 + o(1)).
\eeq
Note that~\req{zEquationAs} is similar to~\req{zEquation}. Finally, using \req{zgAs}~and~\req{zEquationAs} one can obtain the equalities~\req{QUVAsL}. And we get the formula~\req{TAsL} again. $\quad \square$


\newpage



\begin{thebibliography}{99}

\bibitem{s71}
\textit{Sobel M.}, \textit{Kumar S.}, \textit{Blumenthal S.},
Symmetric Binomial Group-Testing with Three Outcomes,
\textit{Purdue Symposium on Statistical Decision Theory and Related Topics}, 1971.

\bibitem{dr83}
\textit{D'yachkov A.G.},  \textit{Rykov V.V.},
A Survey of Superimposed Code Theory,
\textit{Problems of Control and Inform. Theory}, vol.~12, no.~4, pp.~229-242, 1983.

\bibitem{ks64}
\textit{Kautz W.H.}, \textit{Singleton R.C.},
Nonrandom Binary Superimposed Codes,
\textit{IEEE Trans.  Inform. Theory}, vol.~10, no.~4, pp.~363-377, 1964.

\bibitem{dr81}
\textit{D'yachkov A.G.},  \textit{Rykov V.V.},
An Application of Codes for the  Multiple Access Channel in the  ALOHA Communication System,
\textit{Proccedings of the 6-th All-Union Seminar in Computing Networks, Moscow-Vinnitsa}, vol.~4, pp.~18-24, 1981 (in Russian).

\bibitem{d98}
\textit{D'yachkov A.G.}, \textit{Rykov V.V.},
Superimposed Codes for Multiple Accessing of the OR-channel,
\textit{1998 IEEE International Symposium on Information Theory, Boston, USA}, Aug. 1998.

\bibitem{d03}
\textit{D'yachkov A.G.}
Lectures on Designing Screening Experiments,
\textit{Lecture Note Series~10}, Combinatorial and Computational Mathematics Center, Pohang  University of Science and Technology (POSTECH), Korea Republic, Feb. 2003 (survey, 112 pages).

\bibitem{fgu69}
\textit{Friedman A.D.}, \textit{Graham R.L.}, \textit{Ullman J.D.},
Universal single transition time asynchronous state assignments,
\textit{IEEE Trans. Comput. }, vol.~18, no.~6, pp.~541-547, 1969.

\bibitem{cs03}
\textit{Cohen G.D.}, \textit{Schaathun H.G.},
Asymptotic overview on separating codes,
\textit{Tech. Report 248}, Department of Informatics, University of Bergen, Bergen, Norway, 2003.

%\bibitem{m11}
%\textit{Emad A.}, \textit{Shen J.}, \textit{Milenkovic O.},
%Symmetric Group Testing and Superimposed Codes,
%\textit{Proc. IEEE Inf. Theory Workshop (ITW’11)}, pp.~20-24, Oct. 2011.

\bibitem{v98}
\textit{Vilenkin P.A.},
On Constructions of List-Decoding Superimposed Codes,
\textit{Proc. 6th Int. Workshop on Algebraic and Combinatorial Coding Theory (ACCT-6), Pskov, Russia}, pp. 228-231, 1998.

\bibitem{sholomov13ran}
\textit{Sholomov L.A.},
Binary Representation of Underdetermined Data,
\textit{Doklady Akademii Nauk}, vol. 448, no. 3, pp. 275-278, 2013.

\bibitem{sholomov13}
\textit{Sholomov L.A.},
Binary Representations of Underdetermined Data and Superimposed Codes,
\textit{Prikl. Diskr. Mat.}, no. 1, pp. 17-33, 2013 (in Russian).

\bibitem{d14}
\textit{D'yachkov A.G.}, \textit{Vorobyev I.V.}, \textit{Polyanskii N.A.},  \textit{Shchukin V.Yu.},
Bounds on the Rate of Disjunctive Codes,
\textit{Problems of Information Transmission}, vol. 50, no. 1, pp. 27-56, 2014.

\bibitem{d14_isit}
\textit{D'yachkov A.G.}, \textit{Vorobyev I.V.}, \textit{Polyanskii N.A.},  \textit{Shchukin V.Yu.},
Bounds on the Rate of Superimposed Codes,
\textit{2014 IEEE International Symposium on Information Theory},
pp. 2341-2345, Honolulu, HI USA, Jun.29-Jul.4, 2014.

\bibitem{r89}
\textit{Ahmed M. Rashad},
On Symmetrical Superimposed Codes,
\textit{J. Inf. Process. Cybern EIK 29}, vol.~7, pp.~337-341, 1989.

\bibitem{drr89}
\textit{D'yachkov A.G.},  \textit{Rykov V.V.}, \textit{Rashad A.M.},
Superimposed Distance Codes,
\textit{Problems of Control and Inform. Theory}, vol.~18, no~4, pp.~237-250, 1989.

%\bibitem{d14ad}
%\textit{D'yachkov A.G.}, \textit{Vorobyev I.V.}, \textit{Polyanskii N.A.},  \textit{Shchukin V.Yu.},
%Almost Disjunctive List-Decoding Codes,
%\textit{arXiv}:1407.2482 [cs.IT], 2014.

\bibitem{ck81}
\textit{Csiszar I.},  \textit{Korner J.},
Information Theory: Coding Theorems for Discrete Memoryless Systems.
Akademiai Kiado, Budapest, 1981.

\bibitem{opu}
\textit{Galeev E.M.}, \textit{Tikhomirov V.M.},
Optimization: theory, examples, problems.
Editorial URSS, Moscow, 2000. (in Russian)

\end{thebibliography}
\end{document}